\documentclass[12 pt]{iopart}

\usepackage{graphicx}
\usepackage{dcolumn}
\usepackage{bm}
\newcommand{\pderiv}[2]{\frac{\partial #1}{\partial #2}}
\newcommand{\pb}{p_{\rm b}}
\newcommand{\pd}{p_{\rm d}}
\newcommand{\pdc}{p_{\rm {d_c}}}
\newcommand{\rhobar}{\bar\rho}
\newcommand{\ccc}[0]{\circ\circ\circ}
\newcommand{\bcc}[0]{\bullet\circ\circ}
\newcommand{\cbc}[0]{\circ\bullet\circ}

\newcommand{\bbc}[0]{\bullet\bullet\circ}

\newcommand{\bcb}[0]{\bullet\circ\bullet}
\newcommand{\bbb}[0]{\bullet\bullet\bullet}
\newcommand{\bb}[0]{\bullet\bullet}
\newcommand{\bc}[0]{\bullet\circ}
\newcommand{\cb}[0]{\circ\bullet}
\newcommand{\cc}[0]{\circ\circ}
\newcommand{\deriv}[2]{\frac{{\rm d}#1}{{\rm d}#2}}

\begin{document}

\title{Accuracy of the cluster-approximation method
in a nonequilibrium model}

\author{Alastair Windus and Henrik Jeldtoft Jensen}
\address{Department of Mathematics, Imperial College London,
South Kensington Campus, London SW7 2AZ. \\ The Institute for Mathematical Sciences, 53 Prince's Gate, South
Kensington, London SW7 2PG.}
\ead{h.jensen@imperial.ac.uk}
\begin{abstract}
We examine a model in which a nonequilibrium phase transition from an active to an extinct state is observed. The order of this phase transition has
been shown to be either continuous or first-order, depending on the parameter
values and the dimension of the system. Using increasingly large clusters,
we use the cluster approximation method to obtain estimates for the
critical points in 1+1 dimensions. For the continuous phase transitions
only, extrapolations of these approximations show excellent agreement with
simulation results. Further, the approximations suggest that, consistent with simulation results, in 1+1 dimensions no first-order phase transitions are observed.

\end{abstract}


\section{Introduction}
The order of phase transitions in 1+1 dimensions has long been
a topic of debate among researches in the field of nonequilibrium phase transitions.
While it has been argued that first-order phase transitions are impossible
in this dimension (see for example \cite{Hinrichsen_First-order}), most agree that such transitions are feasible. For example, Dickman and Tom\'e
sought to find the simplest model with short-range interactions that exhibited a first-order phase transition in one spatial dimension  \cite{Dickman_1st}. They examined the pair- and triplet-creation models with the reactions
        \begin{equation}
        nA\longrightarrow (n+1)A \quad\mbox{and}\quad A\longrightarrow \phi         \end{equation}
with $n=2$, 3 respectively. The mean field (MF) of such reactions yields a first-order phase transition, yet often, continuous transitions are thought
to be observed in (1+1)-dimensional monte carlo (MC) simulations. They varied the diffusion rate $D$ to see what effect this had since they expected that, with larger diffusion, the model would exhibit more MF-like behaviour due the better mixing of particles, and hence, a first-order phase transition might be observed. They found a first-order phase transition for the triplet-creation
model only. A continuous phase transition was observed in the pair-creation
model even when 95\% of the attempted moves were diffusive. For the triplet-creation
model, hysteresis was observed, indicative of a first-order transition, for
sufficiently high diffusion rate. Recently, Fiore and de Oliveira
\cite{Fiore} used the conservative diffusive contact process, in which the number of particles is fixed, to confirm Dickman and Tom\'e's original findings
(see also \cite{Tome, Cardozo}).
Cardozo and Fontanari \cite{Cardozo} also examined the triplet-creation model and found that the critical exponents change continuously from DP values to compact directed percolation values as the diffusion rate was increased. Due to the strong crossover effects, however, they were unable to locate the precise position of the point where the transition changed order.

In this paper, we wish to examine a slightly modified version of the pair-creation
model in the (1+1)-dimensional case through both simulation and, importantly, MF techniques. The model has previously been shown, by MF, to exhibit both continuous and first-order phase transitions whose lines in parameter space
meet at a tricritical point \cite{Windus_Cluster} (see also \cite{Windus,Grassberger_Tricritical,Lubeck_Tricritical}). In (1+1)-dimensional MC simulations, however, the model exhibits a continuous phase transition across the whole phase space. To examine this model further, we employ a technique originally introduced by ben-Avraham and K\"ohler \cite{ben-Avraham} which they called the $n$-site cluster approximation. In most cases, as $n$ increases, the method
predicts increasingly accurate behaviour of the model in question. In this
paper, we collect data for $n\leq5$ and extrapolate our findings as $n\rightarrow\infty$
to obtain approximations for the order of the phase transition and value
of the critical point. 

\section{The model}  \label{S: The model}
We have a $d$-dimensional square lattice of linear length $L$ where each site is either
occupied by a single particle or is empty. A site is chosen at random. The particle on an occupied site dies with probability $p_{\rm d}$,  leaving the site empty. If the particle does not die, a nearest neighbour site is randomly chosen. If the neighbouring site is empty the particle moves there and produces a new individual at the site that it has just left with probability $k$. If the chosen site is, however,
occupied the particle reproduces with probability $p_{\rm b}$ producing a new particle on another randomly selected neighbouring site, conditional on that site being empty. A time step is defined as the number of lattice sites $N=L^{\rm d}$ and periodic boundary conditions are used. 

We have the following reactions for a particle $A$ for proliferation and
annihilation respectively,
        \begin{equation} \label{E: Reactions}
        A+A+\phi\longrightarrow3A, \quad A+\phi\longrightarrow 2A \quad \mbox{and}         \quad A\longrightarrow \phi.
        \end{equation}

\begin{figure}[tb]
        \centering\noindent
        \includegraphics[width=8.5cm]{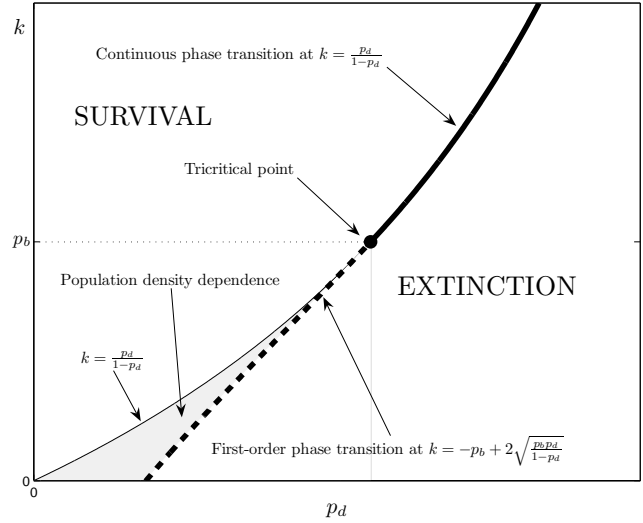}
        \caption{Mean field phase diagram showing the continuous phase transition         occurring
        for $k \ge\pb$ and the first-order phase transition for $k<\pb$.
        Equations (\ref{E: Critical Point 1}-\ref{E: Critical Point 2}) have         been re-arranged in the figure to make $k$ the dependent variable.}
        \label{F: Phase Diagram}
        \end{figure}

Assuming the particles are spaced homogeneously, the MF equation for the density of active sites $\rho(t)$ is given by
        \begin{equation} \label{E: Mean Field}
        \deriv{\rho(t)}{t}  =  \pb\left(1-\pd\right)\rho(t)^2\left(1-\rho(t)\right)\nonumber\\
         +  k(1-\pd)\rho(t)\left(1-\rho(t)\right)-\pd\rho(t).
        \end{equation}
The first two terms consider the sexual and asexual reproduction reactions
respectively and the final term death of an individual. Equation (\ref{E: Mean
Field}) has three stationary states:
        \begin{eqnarray} \label{E: Steady States}
        \bar\rho_0 & = & 0,\\
        \bar\rho_\pm & =& \frac{1}{2}
        \left[1-\frac{k}{\pb}\pm\sqrt{\left(\frac{k}{\pb}-1\right)^2
        +\frac{4}{\pb}\left(k-\frac{\pd}{1-\pd}\right)}\right]. 
        \end{eqnarray}  
For $k \ge \pb$, $\rhobar_+\rightarrow 0$ continuously as $\pd \rightarrow k/(1+k)$, indicative of a continuous phase transition with critical point
        \begin{equation} \label{E: Critical Point 1}
        \pdc(k\ge\pb) = \frac{k}{1+k}.
        \end{equation}
For $k < \pb$, we have a jump in $\rhobar_\pm$ from $(\pb-k)/2\pb$ to
zero, this time at the critical point
        \begin{equation} \label{E: Critical Point 2}
        \pdc(k<\pb) = \frac{(k+\pb)^2}{4\pb +(k+\pb)^2}.        
        \end{equation}
Further, for $k < \pb$, we have a region 
        \begin{equation}
        \frac{k}{1+k} \;<\; \pd \;\leq\; \frac{(k+\pb)^2}{4\pb +(k+\pb)^2}
        \end{equation}        
where the survival of the population is dependent on the population density. In fact, we have extinction for
        \begin{equation}
        \rho(t) < \rhobar_-(k < \pb, \pd).
        \end{equation} 
For $k < \pb$ we therefore have a first-order phase transition. The two phase
transition lines meet at the point $k = \pb$, defining the position of
the tricritical point $k^*$. At the MF level then, we have a phase
diagram as shown in figure \ref{F: Phase Diagram}.
In the region to the left of the transition lines, there exists at least
one real and positive steady state. In the shaded region only, there exist two such steady states with none existing to the right of the transition
lines. The tricritical point not only marks the intersection of the first-order
and continuous phase lines, but also the line bordering the region with population density dependence.  

\section{The cluster approximation technique}
The MF technique that we employed in the previous section assumed independence
between individual sites. In reality there exist, of course, correlations
between nearby sites. Increasing improvements to our original MF equation could then
be made by considering pairs, triplets, quadruplets and so on, of adjacent sites. Such an approach was developed by ben-Avraham and K\"ohler, called
the $(n,m)$-cluster approximation method \cite{ben-Avraham}.

We direct the interested reader to ben-Avraham and K\"ohler's paper \cite{ben-Avraham}
and others (for example \cite{Odor_Cluster,Szolnoki} and references therein)
for a detailed explanation of the approach. Briefly, the method involves examining clusters of size $n$. For sites $1,2,\dots,j$,
we denote $P_{s_1s_2\dots s_j}(t)$ as the probability that the $j$ sites
are in state $\{s_1s_2\dots s_j\}$ at time $t$. The method is concerned with how $P_{s_1s_2\dots s_n}(t)$ changes in time and involves deriving a system of master equations for $P_{\{s_i\}}(t)$. Since $s_i=0$,1 in our case,
there are $2^n$ such equations. It is, however, easy to see that the number
of independent equations is much smaller than this (see, for example, \cite{ben-Avraham}).

Since we are concerned with clusters of size $n$, we need a way to approximate
$P_{s_1s_2\dots s_n}$ from clusters of size $N>n$. In their $(n,m)$-approximation, ben-Avraham and K\"ohler consider adjacent clusters of size $n$ with an overlap
of $m<n$ sites. Using the Bayesian extension process, the approximation is
then,
        \begin{equation}
        P_{s_1s_2\dots s_N} = \frac{\prod_{j=0}^{N-n}P_{s_{j+1}\dots s_{j+n}}}
        {\prod_{j=n-m}^{N-n}P_{s_{j+1}\dots s_{j+m}}}.
        \end{equation}
For example, the $(3,2)$ approximation for a cluster of size six is given by
        \begin{equation}
        P_{s_1s_2s_3s_4s_5s_6} = P_{s_1s_2s_3}\frac{P_{s_2s_3s_4}}{P_{s_2s_3}}
        \frac{P_{s_3s_4s_5}}{P_{s_3s_4}}\frac{P_{s_4s_5s_6}}{P_{s_4s_5}}.
        \end{equation}
Ben-Avraham and K\"ohler found that the $(n,n-1)$-approximation yields the
most accurate results and is termed the $n$-site approximation for short.

\subsection{The 2-site approximation} 
The simple 1-site approximation is just our original MF equations, so we
examine the 2-site approximation. Introducing the subscripts $\bullet$ and
$\circ$ for occupied and empty sites respectively, we have two independent variables, chosen to be the particle density $\rho(t) = P_\bullet(t)$ and the pair density $c(t)=P_{\bb}(t)$. It is then simple to derive the other
2-site probabilities
        \begin{eqnarray}
        d(t) & = & P_{\bc}(t)  =  \rho(t)-c(t) = P_{\cb}(t), \\
        e(t) & = & P_{\cc}(t) = 1-2\rho(t)+c(t).
        \end{eqnarray}

To obtain the master equations, we consider the reactions which change the number of occupied sites $n_\bullet$ and the number of pairs of sites $n_{\bb}$.
The reactions are listed in table \ref{T: Pair approximation}
along with their probabilities of occurring, given the particle configuration.
From this table, the master equations for $\rho$ and $c$ are then derived in the usual way.
        \begin{table}[t]
        \centering 
        \begin{tabular}{cccrrlr} 
        \hline\hline
        \multicolumn{3}{c}{\textbf{Reaction}} & $\Delta n_\bullet$ & $\Delta
        n_{\bullet\bullet}$
        & \textbf{Probability} \\
        \hline\hline
        $\bbb$ & $\longrightarrow$ & $\bcb$ & $-1$ & $-2$ & $\pd c^2/\rho$
        \\
        $\bbc$ & $\longrightarrow$ & $\bcc$ & $-1$ & $-1$ & $\pd cd/\rho$
        & $\times 2$ \\
        $\cbc$ & $\longrightarrow$ & $\ccc$ & $-1$ & $0$ & $\pd d^2/\rho$
        \\ \hline 
        $\bb\cc$ & $\longrightarrow$ & $\bc\bc$ & $0$ & $-1$ & $\frac{1}{2}(1-\pd)(1-k)
        cde/\rho(1-\rho)$ & $\times 2$  \\
        $\cb\cb$ & $\longrightarrow$ & $\cc\bb$ & $0$ & $+1$ & $\frac{1}{2}(1-\pd)(1-k)
        d^3/\rho(1-\rho)$ & $\times 2$  \\
        \hline
        $\bcc$ & $\longrightarrow$ & $\bbc$ & $+1$ & $+1$ & $\frac{1}{2}(1-\pd)k
        de/(1-\rho)$ & $\times 2$  \\
        $\bcb$ & $\longrightarrow$ & $\bbb$ & $+1$ & $+2$ & $\frac{1}{2}(1-\pd)k
        d^2/(1-\rho)$ & $\times 2$  \\
        \hline
        $\bb\cc$ & $\longrightarrow$ & $\bb\bc$ & $+1$ & $+1$ & $\frac{1}{2}(1-\pd)\pb
        cde/\rho(1-\rho)$ & $\times 2$  \\
        $\bb\cb$ & $\longrightarrow$ & $\bb\bb$ & $+1$ & $+2$ & $\frac{1}{2}(1-\pd)\pb
        cd^2/\rho(1-\rho)$ & $\times 2$  \\
        \hline\hline
        \end{tabular}
        \caption{Reactions for the 2-site approximation where, reactions
        such as $\bb\cb\longrightarrow\bc\bb$ for which $\Delta n_\bullet
        = \Delta n_{\bullet\bullet}=0$, have been ignored. A symmetry factor
        arising from the parity symmetry has been included (right column)
        rather than writing both equations down.}
        \label{T: Pair approximation}
        \end{table} 

We note that whereas diffusion of the particles did not feature at all in
the 1-site approximation, it does appear in the 2-site approximation since
$n_{\bb}$ can be affected (see rows four and five in table \ref{T: Pair approximation}).

\begin{figure}[tb]
        \centering\noindent
        \includegraphics[width=10cm]{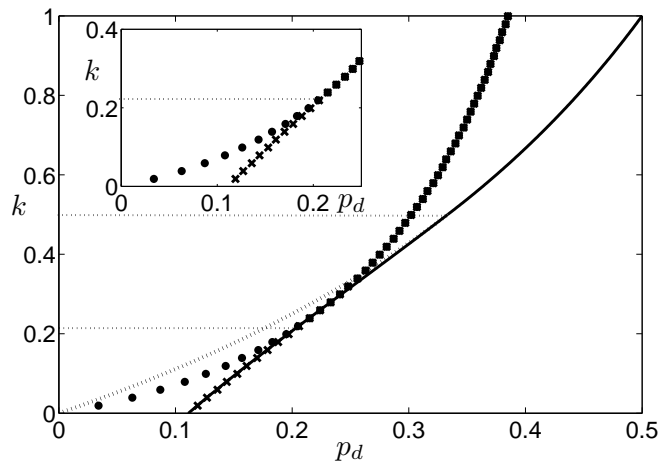}  
        \caption{Phase diagram according to the cluster approximation method.
        The lines show the original MF and the markers the numerically obtained
        values for the 2-site approximation. The region in between the two
        lines and between the different markers show the density dependence
        region. The inset shows the intersection of the two markers for the
        2-site approximation showing the tricritical point. The horizontal
        hashed lines in both plots show the values of $k^*$ according to
        both approximations.}
        \label{F: Site and Pair approximations}
        \end{figure}

Deriving the master equations for $\rho$ and $c$, we have
        \begin{eqnarray}
        \deriv{\rho}{t} & = & (1-\pd)\frac{(\rho-c)(\pb c+k\rho)}{\rho} -
        \pd\rho,         \\
        \deriv{c}{t} & = & (1-\pd)\frac{(\rho-c)\left[(1-k)(\rho^2-c)+(1-c)(k\rho+\pb
        c)\right]}{\rho(1-\rho)} -  2\pd c,         
        \end{eqnarray}        
where we have used such relations as $d+e=1-\rho$. We notice that, if we
make the assumption that all the sites are independent so that $c=\rho^2$,
we return to our original MF equation. Solving the $\rho$-equation, we have
the steady states
        \begin{eqnarray}
        \rhobar_0 & = & 0,\nonumber \\
        \rhobar_\pm & = & \frac{c\left[
        \pb-k\pm\sqrt{(\pb-k)^2-4\pb\left(\frac{\pd}{1-\pd}-k\right)}\right]}{2\left(\frac{\pd}{1-\pd}-k\right)}.
        \end{eqnarray}
These roots are very similar to the roots found in the original MF approximation
except for the extra prefactor 
        \begin{equation}
        \frac{c}{\left(\frac{\pd}{1-\pd}-k\right)}.
        \end{equation} 
Similar to before then, so long as $c>0$, for $k<\pb$, we have a first order phase transition
at 
        \begin{equation}
        \pd = \frac{(k+\pb)^2}{4\pb +(k+\pb)^2}.
        \end{equation} 
We note that critical point $\pdc=k/(1+k)$ is no longer valid
because of the denominator in the prefactor. To find the critical points
then, we solve these equations numerically and plot
the results for the 1-site and 2-site approximations in figure \ref{F: Site and Pair approximations}.

To determine the position of the critical points, for values of $k$
and $\pd$, we numerically found all of the real steady states with $0<\bar c, \rhobar
\leq 1$. Counting the number of such steady states indicated which region
of the phase diagram we were in. Using an iterative procedure enabled us
to locate the boundaries between the regions with zero, one and two such
steady
states as outlined in the phase diagram in figure \ref{F: Phase Diagram}. For increasing $\pd$, the first-order
transition lines were indicated by the number of such steady states changing
from two to zero and from one to zero for the continuous transition. The
tricritical point $k^*$ is then given by the intersection of these two lines.

We can further test the stability of the steady states by examining the
stability matrix
        \begin{equation}
        A = \left(
        \begin{array}{cc}
        \pderiv{}{\rho}\left(\deriv{\rho}{t}\right) & \pderiv{}{c}\left(\deriv{\rho}{t}\right) \\
        \pderiv{}{\rho}\left(\deriv{c}{t}\right) & \pderiv{}{c}\left(\deriv{c}{t}\right)
        \end{array}
        \right),
        \end{equation}
evaluated at the steady states. By calculating the eigenvalues of $A$,
we have that if both eigenvalues are negative, the steady state
is stable, otherwise it is unstable. Testing the non-zero steady states, we find an unstable state in the density dependent region only.

As we see in figure \ref{F: Site and Pair approximations}, for small values
of $k$, the analytical value for the critical
points from the 1-site and 2-site approximations are identical.
We see, further, that the position of the
tricritical point (intersection of the different markers in the figure) is at a lower value of $k$ than the 1-site approximation. Numerically, we found it to be at $k=0.2139$ - less than half of the original MF prediction.

\subsection{Higher order approximations} \label{S: CA Results}

Until now, isolated particles have not been considered appropriately. If
we aim to build a more accurate approximation, we need to be able to consider such
 particles. By examining the 3-site approximation, for example,
we consider such probabilities as $P_{\cbc}$. Clearly, the price that we pay is an increase in complexity since the number of variables and equations increase rapidly. We derived equations for higher order approximations for $n\leq5$. The
$n=5$ case required 13 independent variables with over 1,100
reactions having to be considered. 

\begin{figure}[tb]
        \centering\noindent
        \begin{tabular}{cc}
        {\small a)}  & {\small b)} \\
        \includegraphics[width=7.9cm]{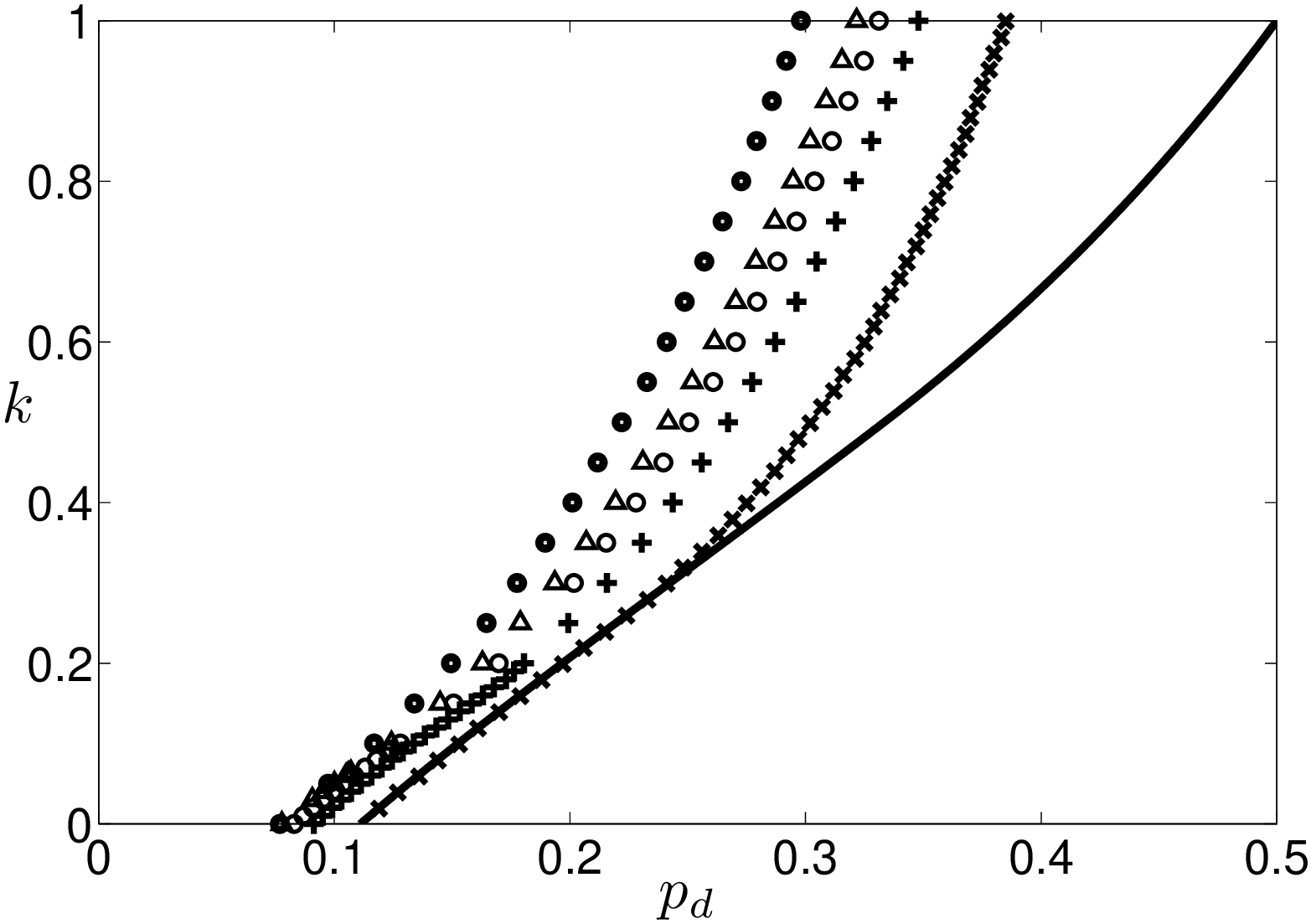} &
        \includegraphics[width=7.9cm]{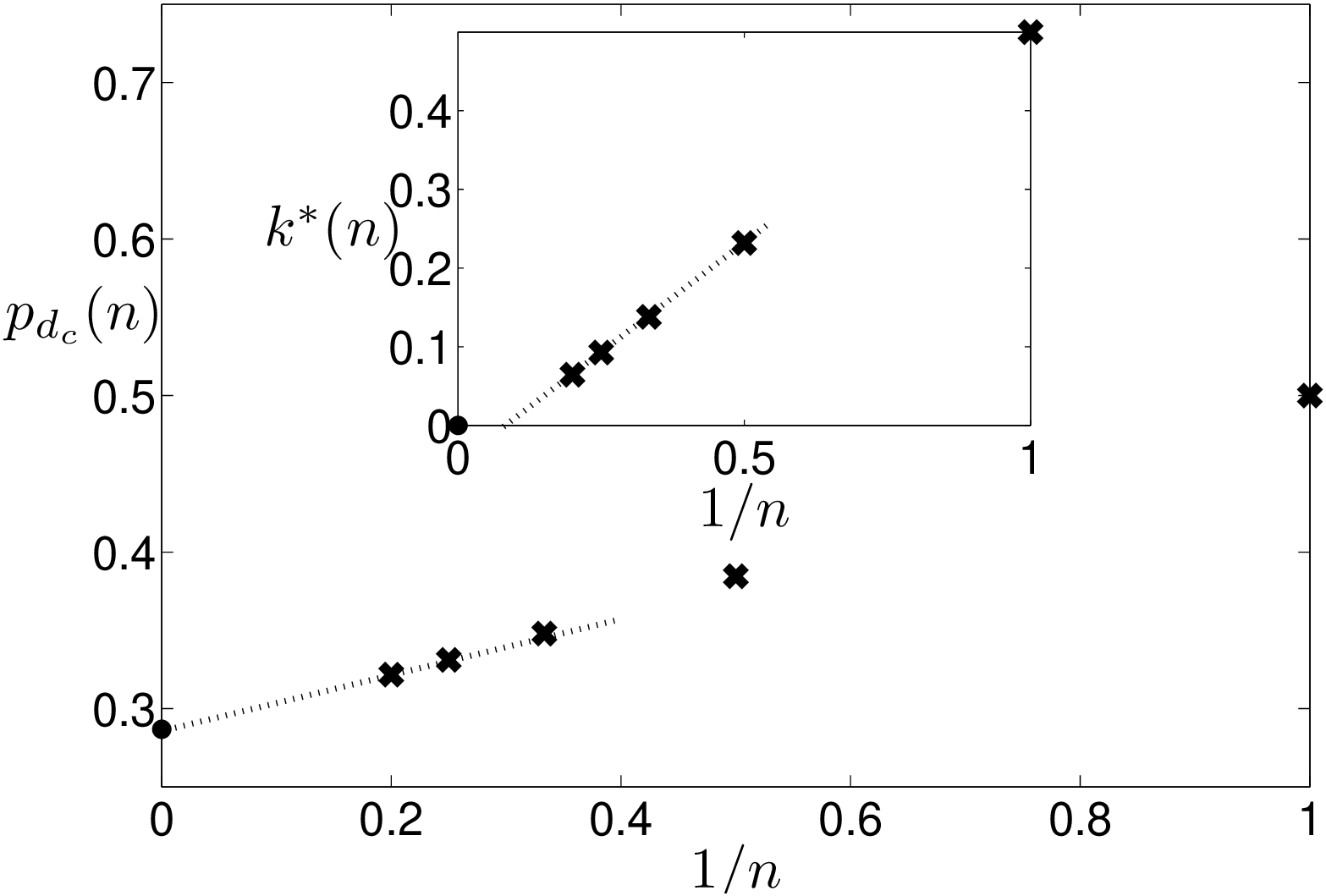}
        \end{tabular}
        \caption{a) Numerical results for the critical point for various         values of $k$. The line shows the original MF approximation
        ($n=1$ and
        the markers (from right to left) the $n=2$, 3, 4 and 5. The     
        circles show the numerical simulation results. c) The approximation
        for $\pdc$ for $k=1$ for the different values of $n$. The circle
        shows the numerical value with the hashed line showing an extrapolation
        through the points for $n=4$ and $n=5$. The inset shows the tricritical
        point $k^*(n)$ with, again, a hashed line showing an extrapolation.}
        \label{F: CA Phase Diagram}
        \end{figure}

The results for the approximations and simulation results for the critical
points are shown in figure \ref{F: CA Phase Diagram} a).
We clearly see how increasing the size of the clusters gives more accurate
approximations for the behaviour of the model when comparing to the MC simulation
results. In particular, for $k=1$, we show in the main plot of figure \ref{F: CA Phase Diagram}
b), the approximated values for $\pdc(n)$ against $1/n$. An extrapolation of the results as $n\rightarrow\infty$ shows excellent agreement
with the MC value. Unfortunately, at the first-order phase transition, when
the number of real and positive steady state solutions decrease from two to zero, such
an extrapolation does not lead to good agreement. This is likely to be due
to the fact that at the continuous phase transition, the correlation length
is infinite and therefore considering increasingly large numbers of adjacent
sites will lead to more accurate approximations. At first-order phase
transitions, since the correlation length remains finite, for $n>\xi_\perp$
we would expect the approximations for $\pdc$ to be independent of $n$.

We can further plot the value of the tricritical point as a function of cluster
size $n$ and, again, extrapolate. As the inset of figure \ref{F: CA Phase Diagram} b) shows, the position of $k^*(n)$ decreases with $n$, appearing to become
zero for some finite $n$. In other words, for sufficiently large $n$, the
corresponding $n$-site cluster approximation would predict no first-order
phase transition. This, however, would be surprising since we would expect
such behaviour to be true only as $n\rightarrow\infty$.  Since the tricritical transition is  infinitesimally close to a first-order transition though, we may well expect a tailing-off of this apparent linear behaviour for large
$n$.

\section{Conclusions}
We have seen how, for this model, the analytical cluster-approximation method seems to correctly predict a continuous phase transition in 1+1 dimensions across
the whole phase space as $n\rightarrow\infty$. Further, the method predicts,
with a high degree of accuracy, the critical point for continuous phase transitions only. These findings highlight the power of the method for this case and
will hopefully lead on to further examination of the techniques involved. 
\ack
All computer simulations were carried out on the Imperial College London's HPC for which we thank Matt Harvey and Simon Burbidge. Alastair Windus would also like to thank EPSRC for his Ph.D. studentship.

\section*{References}
\providecommand{\newblock}{}

\end{document}